\documentclass[twocolumn,superscriptaddress,amsmath, amssymb, amsfonts,preprintnumbers,aps,prd,longbibliography,nofootinbib]{revtex4-1}

\usepackage{scalerel}
\usepackage{color}
\usepackage{amsmath,amsfonts,amssymb}
\usepackage[small,bf,hang]{caption}
\usepackage{slashed}
\usepackage{latexsym,epsfig}
\usepackage{dsfont}
\usepackage{arydshln}
\usepackage{extarrows}
\usepackage{hyperref}

\newcommand{\sfrac}[2]{{\textstyle\frac{#1}{#2}}}

\def\moth{\mathsurround=0pt}
\newdimen\zo \zo=0pt

\def\tick{\leaders\hrule height 0.5ex depth 0pt \hskip 0.5pt}
\def\upboxfill{$\moth \setbox\zo\hbox{\tick}%
  \hskip 3pt\hbox to 0pt{$\tick$\hss}\hrulefill \hbox to 7.5pt{$\tick$\hss}$}

\def\dtick{\leaders\hrule height .34pt depth 0.5ex \hskip 0.5pt}
\def\downboxfill{$\moth \setbox\zo\hbox{\dtick}%
  \hskip 2pt\hbox to 0pt{$\dtick$\hss}\hrulefill \hbox to 2pt{$\dtick$\hss}$}

\def\ov{\bar}

\def\bec{\begin{center}}
\def\ec{\end{center}}

\def\d{\delta}

\def\vf{\varphi}

\def\m{\mu}
\def\n{\nu}
\def\r{\rho}

\def\y{\eta}

\def\nn{\nonumber}

\def\be{\begin{equation}}
\def\ee{\end{equation}}
\def\bea{\begin{eqnarray}}
\def\eea{\end{eqnarray}}
\def\ba{\begin{array}}
\def\ea{\end{array}}

\def\gYM{g_{\text{YM}}}

\begin{document}

\preprint{
HU-EP-21/27-RTG
}
\title{Double Field Theory as the Double Copy of Yang-Mills}

\author{Felipe D\'iaz-Jaramillo} 
\email{felipe.diaz-jaramillo@hu-berlin.de}
\affiliation{%
Institut f\"ur Physik und IRIS Adlershof, Humboldt-Universit\"at zu Berlin,
Zum Gro{\ss}en Windkanal 2, 12489 Berlin, Germany
}

\author{Olaf Hohm} 
\email{ohohm@physik.hu-berlin.de}
\affiliation{%
Institut f\"ur Physik und IRIS Adlershof, Humboldt-Universit\"at zu Berlin,
Zum Gro{\ss}en Windkanal 2, 12489 Berlin, Germany
}

\author{Jan Plefka} 
\email{jan.plefka@hu-berlin.de}
\affiliation{%
Institut f\"ur Physik und IRIS Adlershof, Humboldt-Universit\"at zu Berlin,
Zum Gro{\ss}en Windkanal 2, 12489 Berlin, Germany
}


\begin{abstract}
We show that double field theory  arises from the color-kinematic double copy of Yang-Mills theory.
A precise double copy prescription for the Yang-Mills action 
at quadratic and cubic order is provided that yields the double field theory action 
in which the duality invariant dilaton has been integrated out. 
More precisely, at quadratic order this yields  the gauge invariant double field 
theory, while at cubic order it yields 
the cubic double field theory action subject to a gauge condition that originates 
from Siegel gauge in string field theory.

\end{abstract}

\maketitle

\section{Introduction}

The fundamental interactions in nature, as far as currently known,  are governed by two kinds of theories: 
Yang-Mills theory, which describes the gauge bosons of the standard model of particle physics, 
and Einstein's theory of general relativity, which describes the force of gravity and the universe at large. 
While these two kinds of theories share broad qualitative features, such as the presence of gauge symmetries, 
their actions appear to be  quite  different. Most importantly, at the quantum field theory level, the perturbative expansion of general relativity about flat space-time is 
notoriously involved and non-renormalizable, in sharp contrast to the well-understood perturbation theory of Yang-Mills theory. 

Yet, the Bern-Carrasco-Johansson construction \cite{Bern:2008qj,Bern:2010ue} of
 gravity scattering amplitudes as the  `double copy' of Yang-Mills amplitudes
suggests   that there might be a much more intimate relationship between gauge theory and gravity. 
Specifically, the perturbation theory of Yang-Mills and more general gauge theories  can be arranged in such a way that their building blocks obey
a property known as color-kinematics duality. Here the contributions associated to the gauge group known as 
`color factors' appear on the same footing as the purely kinematical numerator factors (functions of momenta and
polarizations), which in turn obey relations akin to Jacobi identities. 
Replacing  the color factors by kinematic factors then yields  gravity amplitudes, whose computation by standard textbook 
methods starting from the Einstein-Hilbert Lagrangian is incomparably  more involved, see \cite{Bern:2019prr} 
for a recent review and \cite{quantamagazine} for a popular account. Somewhat misleadingly, these observations are 
often summarized by ``gravity is Yang-Mills squared". Formal proofs of the double copy have been provided at tree-level
\cite{Bern:2010yg,Mafra:2011kj,*Bjerrum-Bohr:2016axv,*Du:2017kpo,*Bridges:2019siz} using a variety of methods. Evidence for the existence also at loop-level has been provided
through a number of explicit constructions, see \cite{Bern:2019prr} for an account
and \cite{Borsten:2020zgj,Borsten:2021hua} for a treatment in terms of homotopy algebras. 
(In addition, the double copy
prescription has been extended to the sector of classical solutions \cite{Monteiro:2014cda,*Luna:2015paa,*Luna:2016hge,*Carrillo-Gonzalez:2017iyj,*Guevara:2020xjx,*Monteiro:2020plf}.) 

Double copy techniques are  a main feature  of the modern amplitude program, which largely abandons 
the textbook approach to quantum field theory with its baggage of off-shell fields and gauge redundancies, see 
\cite{Dixon:1996wi,*Elvang:2013cua,*Henn:2014yza} for reviews. 
As such, there is no reason to expect that features of the double copy can be seen at the level of an action. 
To quote \cite{Nicolai:2010zza}:  ``no amount of fiddling with the Einstein-Hilbert action will reduce it to a square of a Yang-Mills action." 
Nevertheless, there have been numerous attempts to render the double copy at least partially manifest at the level of a Lagrangian \cite{Bern:1999ji,Bern:2010yg,Tolotti:2013caa,*Anastasiou:2018rdx,*Borsten:2020xbt,*Ferrero:2020vww,*Beneke:2021ilf}.

A first step is to rewrite the perturbative expansion of gravity in a way that manifests a left-right index factorization. 
Starting from the Einstein-Hilbert action this requires a series of elaborate field redefinitions  that were worked out to 
a small order in fields in \cite{Bern:1999ji}. 
This problem is   solved in the framework of double field theory \cite{Siegel:1993th,Hull:2009mi,Hull:2009zb,Hohm:2010jy,Hohm:2010pp} that doubles space-time yielding an
automatic index factorization pertaining to the two space-time coordinates.
Double field  theory was originally conceived of as a non-local field theory, referred to as \textit{weakly constrained double field theory}, 
 that captures, by virtue of fields depending on doubled coordinates,   massive Kaluza-Klein and winding modes of string
 theories on 
toroidal backgrounds \cite{Hull:2009mi}.  
While this theory has not yet  been constructed explicitly beyond cubic order,  much work has been done 
on \textit{strongly constrained double field theory}, in which the doubling of coordinates is purely formal
but which provides a duality covariant and background independent 
reformulation of the  target space actions of string theory including metric, B-field and dilaton.  
(See \cite{Aldazabal:2013sca,*Berman:2013eva,*Hohm:2013bwa} for reviews and \cite{Tseytlin:1990va,*Kugo:1992md,*Siegel:1993xq} for earlier important work.) 
It was   anticipated by Siegel  in \cite{Siegel:1993th} that these  reformulations render  index factorizations  manifest, 
and this was established  explicitly to all orders in fields in \cite{Hohm:2011dz} and used in \cite{Boels:2015daa} for amplitude computations. 
(See also \cite{Cheung:2016say,*Cheung:2017kzx}, where the problem of index factorizations was revisited more recently, and \cite{Lee:2018gxc,Kim:2019jwm,Berman:2020xvs} 
for double copy prescriptions of classical solutions in double field theory.)

In this letter we go beyond this  and ask:  what does a double copy prescription applied 
to the Yang-Mills Lagrangian actually give? As the double copy instructs us to replace color factors with a second 
set of kinematic factors, which come with their own  momenta,  this naturally leads to a double field theory with doubled  momenta or, in position space, a 
doubled set of coordinates. The key technical result reported here is that this double copy of Yang-Mills theory yields at quadratic and cubic 
order  double field theory 
upon integrating out the duality invariant dilaton.\footnote{The duality invariant dilaton $\phi$ of double field theory is defined in 
terms of the familiar scalar dilaton $\Phi$ by $e^{-2\phi}=\sqrt{-g}e^{-2\Phi}$.} More precisely, at quadratic order this match holds at the level of gauge invariant Lagrangians, while at cubic order 
this match requires a gauge choice, which in the string field theory formulation of  \cite{Hull:2009mi} originates from the so-called Siegel gauge 
and which reduces  to the de Donder gauge for standard gravity fields.  

\section{Quadratic theory}

We start from the action for Yang-Mills theory in $D$ dimensions, 
 \be\label{actionYM}
  S_{\rm YM} = -\frac{1}{4 }\int d^D x\,\kappa_{ab} \,F^{\mu\nu \,a} F_{\mu\nu}{}^{b}\;, 
 \ee
with  the field strength for the gauge bosons $A_{\mu}{}^{a}$, 
 \be
  F_{\mu\nu}{}^{a} = \partial_{\mu}A_{\nu}{}^{a} -\partial_{\nu}A_{\mu}{}^{a}  +\gYM  f^{a}{}_{bc} A_{\mu}{}^{b} A_{\nu}{}^{c}\;, 
 \ee 
where  $\gYM$ is the gauge coupling.   
Here, $f^{a}{}_{bc}$ denote the structure constants of the color gauge group, with adjoint indices $a,b,\ldots$, 
and invariant Cartan-Killing form $\kappa_{ab}$ that lowers adjoint indices, so that $f_{abc}\equiv \kappa_{ad}f^{d}{}_{bc}$  is totally antisymmetric. 
Expanding the Yang-Mills action to quadratic order in fields and 
integrating by parts one obtains 
 \be
  S_{\rm YM}^{(2)} = \frac{1}{2}\int d^Dx\,\kappa_{ab}\,A^{\mu a} \big(\square A_{\mu}{}^{b} - \partial_{\mu}\partial^{\nu} A_{\nu}{}^{b}\big)\;. 
 \ee
In order to motivate the double copy prescription it is convenient to pass over to momentum space. Defining 
$A_{\mu}{}^{a}(k)\equiv \frac{1}{(2\pi)^{D/2}}\int d^D x A_{\mu}{}^{a}(x)e^{ikx}$ the quadratic action reads 
 \be
  S_{\rm YM}^{(2)} = -\frac{1}{2} \int_{k}\,\kappa_{ab}\,k^2 \,\Pi^{\mu\nu}(k) A_{\mu}{}^{a}(-k) A_{\nu}{}^{b}(k)\; , 
 \ee
where $\int_{k}:=\int d^{D}k$, and we have scaled out $k^2$,   
 in order to define the projector in terms of the Minkowski metric $\eta_{\mu\nu}={\rm diag}(-,+,+,+)$
 \be\label{PiProjector}
  \Pi^{\mu\nu}(k) \equiv  \eta^{\mu\nu}-\frac{k^{\mu}k^{\nu}}{k^2}\;, 
 \ee 
which  obeys the identities  
 \be\label{projIdentity}
  \Pi^{\mu\nu}(k)k_{\nu} \equiv 0\;, \qquad \Pi^{\mu\nu}\Pi_{\nu\rho} \equiv  \Pi^{\mu}{}_{\rho}\;. 
 \ee
 The first identity implies   gauge invariance under 
 \be
   \delta A_{\mu}{}^{a}(k) = k_{\mu} \lambda^{a}(k)  \;, 
 \ee  
where the gauge parameter $\lambda^a(k)$ is an arbitrary function.

Let us  now turn to the double copy construction of a gravity theory. We take the double copy prescription to 
lead to a double field theory: Replace the color indices $a$ 
by a \emph{second} set of spacetime indices denoted by a bar, 
$a\rightarrow \bar{\mu}$, corresponding to a \emph{second} set of spacetime 
momenta $\bar{k}^{\bar{\mu}}$:\footnote{The mass dimensions in momentum
space are $[A]=-1-\frac{D}{2}$ and $[e]=-1-D$, so this substitution does not preserve the dimensions, 
which is due to the additional $\bar{k}$ integral. Similarly, the dimensions of the coupling constants are 
$[\gYM]=2-\frac{D}{2}$ and $[\kappa]=1-D$ for the gravitational coupling  in double field theory, 
which differs from that in Einstein gravity.}  
 \be\label{DcFields}
  A_{\mu}{}^{a}(k) \; \rightarrow \; e_{\mu\bar{\mu}}(k,\bar{k})\;. 
 \ee
To complete the double copy prescription for the quadratic theory we need to define a substitution rule for the Cartan-Killing metric 
$\kappa_{ab}$. We will see that the following  replacement does the job:
 \be\label{DCCartan}
  \kappa_{ab} \; \rightarrow \; \sfrac{1}{2}\, \bar{\Pi}^{\bar{\mu}\bar{\nu}}(\bar{k})\,, 
 \ee
where $\bar{\Pi}^{\bar{\mu}\bar{\nu}}$  is defined as in (\ref{PiProjector}), but with all momenta replaced by barred momenta and all indices replaced by 
barred indices. This prescription is motivated from the double copy rule at the level of amplitudes: for a gauge theory amplitude ${\cal A}=\sum_i \frac{n_ic_i}{D_i}$, 
where $n_i$ are kinematic factors, $c_i$ are color factors, and the $D_i$ are the inverse propagators, the double copy amounts to replacing $c_i$ by 
kinematic factors $n_i$, while the $D_i\sim k^2$ are untouched. Thus, it is natural to scale out $k^2$ from the kinetic operator and to double only 
the resulting projector $\Pi^{\mu\nu}$, which also guarantees that a two-derivative theory is mapped to a two-derivative theory.

The quadratic gravity action following from this double copy (DC) prescription then reads
 \be\label{quadraticDoubleCopy}
  S_{\rm DC}^{(2)} = 
  -\frac{1}{4} \int_{k,\bar{k}} k^2\,   \Pi^{\mu\nu}(k)  \bar{\Pi}^{\bar{\mu}\bar{\nu}}(\bar{k}) e_{\mu\bar{\mu}}(-k,-\bar{k}) 
  e_{\nu\bar{\nu}}(k,\bar{k})\,.  
 \ee
We emphasize that the field $e_{\mu\bar{\mu}}$ now depends on doubled momenta $K\equiv (k,\bar{k})$. 
Note that the momenta $k$ and $\bar{k}$ enter the action on the same footing, except that we have chosen the 
factor in front to be $k^2$ rather than $\bar{k}^2$, but in double field theory this asymmetry is resolved due to the so-called level-matching 
constraint \cite{Hull:2009mi}
 \be\label{levelmatching}
  k^2 = \bar{k}^2\;, 
 \ee 
where two copies of the same flat  space-time metric are used to take the square.  
In order to match with double field theory, and also to lead to a local action, 
we thus have to assume that the doubled momenta are subject to this constraint (which does have more general 
solutions than the trivial $k=\bar{k}$ for which the theory reduces to a standard linearized  gravity theory). 
We also note that, 
owing to the first identity in (\ref{projIdentity}),  the action is manifestly gauge invariant under  
 \be\label{deltagaugee}
  \delta e_{\mu\bar{\nu}} = k_{\mu}\bar{\lambda}_{\bar{\nu}}  + \bar{k}_{\bar{\nu}}\lambda_{\mu} \;, 
 \ee
with two independent gauge parameters $\lambda_{\mu}$ and $\bar{\lambda}_{\bar{\mu}}$ that depend on doubled momenta 
$K\equiv (k,\bar{k})$, subject to (\ref{levelmatching}). 
 
We will now show that (\ref{quadraticDoubleCopy}) is indeed equivalent to (quadratic) double field theory. 
Writing out the projectors with (\ref{PiProjector}) and using the level-matching constraint (\ref{levelmatching}) the action reads 
  \begin{align}\label{YMsquaredConstrained}
     S_{\rm DC}^{(2)} = -\frac{1}{4} \int_{k,\bar{k}} \Big(&  k^2 e^{\mu\bar{\nu}}e_{\mu\bar{\nu}}-k^{\mu} k^{\rho} e_{\mu\bar{\nu}}e_{\rho}{}^{\bar{\nu}}
     - \bar{k}^{\bar{\nu}}\bar{k}^{\bar{\sigma}} e_{\mu\bar{\nu}} e^{\mu}{}_{\bar{\sigma}} \nn\\ &
     +\frac{1}{k^2} k^{\mu}k^{\rho}\bar{k}^{\bar{\nu}}\bar{k}^{\bar{\sigma}} e_{\mu\bar{\nu}} e_{\rho\bar{\sigma}}
     \Big)\,.
  \end{align}
In order to compare with the standard double field theory action we have to Fourier transform to (doubled) position space. 
This is straightforward except for the last term in (\ref{YMsquaredConstrained}), which due to the factor $\frac{1}{k^2}$ would yield a 
non-local term. This problem is resolved by introducing  an auxiliary scalar field $\phi(k,\bar k)$ (the dilaton):
   \begin{align}\label{localDFTquad}
     S_{\rm DC}^{(2)} = -\frac{1}{4} \int_{k,\bar{k}}\Big(& k^2 e^{\mu\bar{\nu}}e_{\mu\bar{\nu}}-k^{\mu} k^{\rho} e_{\mu\bar{\nu}}e_{\rho}{}^{\bar{\nu}}
     - \bar{k}^{\bar{\nu}}\bar{k}^{\bar{\sigma}} e_{\mu\bar{\nu}} e^{\mu}{}_{\bar{\sigma}} \nn\\ &
          -k^2 \phi^2 +2\phi\, k^{\mu}\bar{k}^{\bar{\nu}} e_{\mu\bar{\nu}}
     \Big)\;. 
  \end{align}
Integrating out  $\phi$ by solving its own field equations, 
 \be\label{PhiSolution}
  \phi = \frac{1}{k^2}k^{\mu}\bar{k}^{\bar{\nu}} e_{\mu\bar{\nu}}\;, 
 \ee
and back-substituting into the action we recover the non-local (\ref{YMsquaredConstrained}). 
Alternatively, without integrating out fields, one may redefine the dilaton as 
$\phi \to \phi' =  \phi - \frac{1}{k^2}k^{\mu}\bar{k}^{\bar{\nu}} e_{\mu\bar{\nu}}$,  
which decouples $\phi'$  from $e_{\mu\bar\nu}$. 
The action  (\ref{localDFTquad}) is of course  still gauge invariant, 
with a gauge transformation for $\phi$ that is determined by the variation of (\ref{PhiSolution}): 
 \be\label{deltaGaugephi}
  \delta\phi = k_{\mu} \lambda^{\mu} + \bar{k}_{\bar{\mu}} \bar{\lambda}^{\bar{\mu}}\;, 
 \ee
where we used (\ref{levelmatching}). 
With the action in the form (\ref{localDFTquad})  it is then straightforward  to Fourier transform to a local action in doubled position space: 
  \begin{align}\label{localDFTquadPosition}
     S_{\rm DC}^{(2)} = \frac{1}{4}& \int d^Dx \,d^D\bar{x} \Big( e^{\mu\bar{\nu}}\square e_{\mu\bar{\nu}}+\partial^{\mu}  e_{\mu\bar{\nu}}\,\partial^{\rho}  e_{\rho}{}^{\bar{\nu}} \nn\\ &
     + \bar{\partial}^{\bar{\nu}} e_{\mu\bar{\nu}}\, \bar{\partial}^{\bar{\sigma}}e^{\mu}{}_{\bar{\sigma}}
          - \phi\square \phi +2\phi \partial^{\mu}\bar{\partial}^{\bar{\nu}} e_{\mu\bar{\nu}}
     \Big) ,
  \end{align}
where $\partial_{\mu} =\frac{\partial}{\partial x^{\mu}}$ and $\bar{\partial}_{\bar{\mu}}=\frac{\partial}{\partial \bar{x}^{\bar{\mu}}}$ are the partial derivatives 
corresponding to the coordinates that are dual to $k^{\mu}$ and $\bar{k}^{\bar{\mu}}$ and hence by (\ref{levelmatching}) subject to the 
constraint 
 \be\label{positionlevelmatching}
  \square \equiv \partial^{\mu} \partial_{\mu} = \bar{\partial}^{\bar{\mu}}\bar{\partial}_{\bar{\mu}}\;. 
 \ee
The gauge transformations (\ref{deltagaugee}) and (\ref{deltaGaugephi}) translate in doubled position space to 
 \be
 \begin{split}
  \delta e_{\mu\bar{\nu}} &= \partial_{\mu}\bar{\lambda}_{\bar{\nu}} + \bar{\partial}_{\bar{\nu}}\lambda_{\mu}\;, \\
  \delta \phi &=  \partial_{\mu} \lambda^{\mu} + \bar{\partial}_{\bar{\mu}} \bar{\lambda}^{\bar{\mu}}\;, 
 \end{split}
 \ee
under which (\ref{localDFTquadPosition}) is invariant, modulo the constraint (\ref{positionlevelmatching}). 
The action (\ref{localDFTquadPosition}) defines precisely the standard quadratic double field theory action, 
which upon setting $x=\bar{x}$ is equivalent, up to field redefinitions, 
to the familiar free action for gravity, antisymmetric tensor and dilaton \cite{Hull:2009mi}. \\

\section{Cubic theory}

We now turn to the cubic vertex of  Yang-Mills theory  and extend the double copy construction to 
the cubic action of double field theory. The cubic part of the Yang-Mills action (\ref{actionYM}) reads 
  \be
   S^{(3)}_{\rm YM} = - \gYM\int d^Dx\, f_{abc}\, \partial^{\mu}A^{\nu a} \,A_{\mu}{}^{b}\, A_{\nu}{}^c\;. 
  \ee
Upon Fourier transforming to momentum space this becomes 
  \be
   S^{(3)}_{\rm YM} = \sfrac{i\gYM}{(2\pi)^{{D}/{2}}} \int_{k_1,k_2,k_3 }
 \!\!\!\!\!\!\!\!\!\!\!\!\! \delta(k_1+k_2+k_3) f_{abc} k_1^{\mu} A_{1}^{\nu\,a} A_{2\mu}{}^{b} A_{3 \nu}{}^c\,, 
  \ee
where we use the short-hand notation $A_i \equiv A(k_i)$, and we 
performed the $x$-integration, 
introducing  the delta function. It is convenient to write this more symmetrically as 
  \begin{align}
S^{(3)}_{\rm YM}=-&\sfrac{i\gYM}{6(2\pi)^{{D}/{2}} }\int_{k_1,k_2,k_3 } 
 \!\!\!\!\!\!\!\!\!\!  \d (k_{1}+k_{2}+k_{3})\\ &\qquad  \times f_{abc}\, \Pi^{\m\n\r}(k_{1},k_{2},k_{3}) 
\, A_{1\m}{}^a A_{2\n}{}^b A_{3\r}{}^c\, ,\nn
\end{align}
where we defined 
\begin{equation}\label{3indexPi}
\Pi^{\m\n\r}(k_{1},k_{2},k_{3})\equiv \y^{\m\n}k_{12}^{\r}+\y^{\n\r}k^{\m}_{23}+\y^{\r\m}k^{\n}_{31}\, ,
\end{equation}
with $k_{ij}\equiv k_{i}-k_{j}$.
Note that
this tensor has the anti-symmetry properties required by the structure it multiplies, e.g., 
   $\Pi^{\mu\nu\rho}(k_1,k_2,k_3) = - \Pi^{\nu\mu\rho}(k_2,k_1,k_3)$.

Our task now is to give  the double copy prescription that extends (\ref{DcFields}), (\ref{DCCartan}) to the cubic theory. 
The natural substitution rule is 
\begin{equation}\label{eq:precripcubic}
f_{abc} \; \to \;  \tfrac{i}{4} \,
\ov\Pi^{\ov\m\ov\n\ov\r}(\ov k_{1},\ov k_{2},\ov k_{3})\;, 
\end{equation}
where the factor of $i$ is needed since we relate a theory with one derivative to a theory with two derivatives. 
Together with  $\gYM\rightarrow \frac{1}{2} \kappa$, after which we set $\kappa=1$,  this gives the cubic action 
\begin{align}
&S^{(3)}_{\rm DC}=\sfrac{1}{48(2\pi)^{D/2}}\int  {dK_{1}}{dK_{2}}{dK_{3}}\, \d (K_{1}+K_{2}+K_{3})\nn \\
&\qquad \times  \ov\Pi^{\ov\m\ov\n\ov\r}(\ov k_{1},\ov k_{2},\ov k_{3}) \, \Pi^{\m\n\r}(k_{1},k_{2},k_{3})
\nn \, 
e_{1\,\m\ov\m}\, e_{2\,\n\ov\n}\, e_{3\,\r\ov\r}\, , 
\end{align}
where we use the short-hand notation $e_{i\, \m\ov\m}\equiv e_{\m\ov\m}(K_{i})$, with $K\equiv (k,\bar{k})$ for doubled momenta, 
and  $dK\equiv d^{2D}K$. 
Writing out $\Pi^{\mu\nu\rho}$ and  $\bar{\Pi}^{\bar{\mu}\bar{\nu}\bar{\rho}}$ yields nine terms which, upon relabeling momentum variables  and  indices, reduce to
two terms, and then writing out $k_{ij}=k_i-k_j$ 
 the action becomes
\begin{widetext} 
\begin{align}
S^{(3)}_{\rm DC}&=\sfrac{1}{8(2\pi)^{D/2}}\int  {dK_{1}}{dK_{2}}{dK_{3}}\, \d (K_{1}+K_{2}+K_{3})\\
&\times\, e_{1\, \m\ov\m}\Big[-k_{2}^{\m}\, e_{2\, \r\ov\r}\,\ov k^{\ov\m}_{3}\, e_{3}^{\r\ov\r}+k^{\m}_{2}\, e_{2\, \n\ov\r}\, \ov k^{\ov\r}_{3}\, e_{3}^{\n\ov\m}+k_{2}^{\r}\, e^{\m\ov\r}_{2}\, \ov k^{\ov\m}_{3}\, e_{3\, \r\ov\r} 
+k_{2}^{\m}\, \ov k_{2}^{\ov\m}\, e_{2\, \r\ov\r}\, e_{3}^{\r\ov\r}-k_{2\, \r}\, e_{2}^{\m\ov\r}\, \ov k_{3\, \ov\r}\, e_{3}^{\r\ov\m}-k_{2}^{\r}\, \ov k^{\ov\m}_{2}\, e_{2}^{\m\ov\r}\, e_{3\r\ov\r}\Big]\;. \nn
\end{align}
Fourier transforming  to position space and integrating by parts, we finally obtain 
\begin{equation}\label{eq:DCBAD}
\begin{split}
S^{(3)}_{\rm DC}=\frac{1}{8}\int d^D x \,d^D\ov x\;  e_{\m\ov\m}\, \Big[& \,2\partial^{\m}e_{\r\ov\r}\,\ov \partial^{\ov\m}e^{\r\ov\r}-2\partial^{\m}e_{\n\ov\r}\,\ov \partial^{\ov\r}e^{\n\ov\m}-2\partial^{\r}e^{\m\ov\r}\,\ov \partial^{\ov\m}e_{\r\ov\r} 
 +\partial^{\r}e_{\r\ov\r}\,\ov \partial^{\ov\r}e^{\m\ov\m}+\ov \partial_{\ov\r}e^{\m\ov\r}\,\partial_{\r}e^{\r\ov\m}\, \Big]\;.
\end{split}
\end{equation}
In the following we will prove that this action agrees precisely with the cubic double field theory upon imposing a gauge fixing condition and integrating out the dilaton. 
The need for a gauge fixing condition is natural in view of amplitude computations: 
The double copy only works in the Feynman respectively de Donder gauges. 
In fact, we have also verified by  a  Noether construction  that 
the sum of the quadratic and cubic double copy actions (\ref{localDFTquadPosition}) 
and (\ref{eq:DCBAD}) cannot be gauge invariant. Thus, at best the above cubic piece (\ref{eq:DCBAD}) is related to a gauge fixed form of double field theory. 

\end{widetext}

It turns out to be convenient to use the form of double field theory originally derived from closed string field theory (SFT) \cite{Hull:2009mi}. 
We will not need any detailed technical background of SFT, but it is instructive to write out the string field truncated to the relevant states. The string field $\Psi$ takes values in 
the first-quantized Hilbert space and reads 
 \begin{align}\label{Psi}
 | \Psi\rangle = \int & dK \Big(-\frac{1}{2} e_{\mu\bar{\nu}}(K) \alpha_{-1}^{\mu} \bar{\alpha}_{-1}^{\bar{\nu}} c_1 \bar{c}_1  \\
  &
  +\varphi(K) c_1 c_{-1} +\bar{\varphi}(K) \bar{c}_{1}\bar{c}_{-1} \nn\\
  & + i f_{\mu}(K) c_0^+ c_1 \alpha^{\mu}_{-1} 
  + i \bar{f}_{\bar{\mu}}(K) c_0^+ \bar{c}_1
  \bar{\alpha}_{-1}^{\bar{\mu}} \Big)|{\bf 0};K \rangle \;, \nn
 \end{align}
where $\alpha^{\mu}_{m}$ and $\bar{\alpha}^{\bar{\mu}}_{m}$ are the familiar left- and right-moving oscillators of the first-quantized string, 
and the $c_m$ and $\bar{c}_m$ are ghosts, being part of a \textit{bc} ghost system with algebra $\{b_m, c_m\}=\delta_{m+n}$ and 
$c_0^{\pm} = \frac{1}{2}(c_0\pm \bar{c}_0)$. The state $|{\bf 0};K \rangle$ is constructed from the so-called $SL(2,\mathbb{C})$-invariant vacuum $|{\bf 0}\rangle$ 
by the action of $\exp(ik\cdot x)$ and $\exp(i\bar{k}\cdot\bar{x})$ and obeys in particular $b_0|{\bf 0};K \rangle = \bar{b}_0|{\bf 0};K \rangle=0$ \cite{Hull:2009mi}, 
see \cite{Arvanitakis:2021ecw}
for a review. 
Closed SFT yields for this truncated string field an action  that reads to cubic order in position space  (see equation (3.3) in \cite{Hull:2009mi})
 \be\label{fullSFT}
  S_{\rm SFT} = \int d^Dx \,d^D\bar{x} \, \Big[{\cal L}^{(2)} + {\cal L}^{(3)}_{\rm DC} + {\cal L}^{(3)}_{\rm extra}  
  \Big]\;, 
 \ee
where the superscripts denote the power of fields, and  the various terms are defined as follows. 
\begin{widetext}
The quadratic action is given by 
 \be\label{quadraticFirstOrder}
  {\cal L}^{(2)} = \sfrac{1}{4}e_{\m\ov\m}\square e^{\m\ov\m}+2\ov\vf\square \vf-f_{\m}f^{\m}-\ov f_{\ov\m}\ov f^{\ov\m}-f^{\m}\left(\ov \partial^{\ov\n}e_{\m\ov\n}
  -2\partial_{\m}\ov\vf\right)+\ov f^{\ov\n}\left(\partial^{\m}e_{\m\ov\n}+2\ov \partial_{\ov\n}\vf\right)\,. 
 \ee 
We observe that $f_{\m}$ and $\ov f_{\ov\m}$ are auxiliary and  can hence be integrated out algebraically. 
Doing so we recover the quadratic double field theory action in the form (\ref{localDFTquadPosition}), 
with $\phi\equiv \varphi-\bar{\varphi}$ and the combination $\varphi+\bar{\varphi}$ dropping out and hence being pure gauge.
The Lagrangian ${\cal L}^{(3)}_{\rm DC}$ precisely agrees with (\ref{eq:DCBAD}) obtained from the double copy construction. 
Thus, the complete double field theory action to cubic order differs from the double copy construction by   ${\cal L}^{(3)}_{\rm extra}$, 
which is given by 
\begin{align}
{\cal L}^{(3)}_{\rm extra}= &\left.  \sfrac{1}{2}e_{\m\ov\n}f^{\m}\ov f^{\ov\n}-\sfrac{1}{2}f_{\m}f^{\m}\ov\vf+\sfrac{1}{2}\ov f_{\ov\n}\ov f^{\ov\n}\vf-\sfrac{1}{4}f^{\m}\left(e_{\m\ov\n}\ov \partial^{\ov\n}\ov\vf+\ov \partial^{\ov\n}\left(e_{\m\ov\n}\ov\vf\right)\right)-\sfrac{1}{4}\ov f^{\ov\n}\left(e_{\m\ov\n}\partial^{\m}\vf+\partial^{\m}\left(e_{\m\ov\n}\vf\right)\right)\right.\\
&\left.+\sfrac{1}{4}f^{\m}\left(\ov\vf \partial_{\m}\vf-\vf \partial_{\m}\ov\vf\right)+\sfrac{1}{4}\ov f^{\ov \n}\left(\ov\vf \ov \partial_{\ov\n}\vf-\vf \ov \partial_{\ov \n}\ov\vf\right)\right.
-\sfrac{1}{8}e_{\m\ov\n}\left(\ov\vf \partial^{\m}\ov \partial^{\ov\n}\vf+\vf \partial^{\m}\ov \partial^{\ov\n}\ov\vf-\partial^{\m}\vf\ov \partial^{\ov\n}\ov\vf
-\ov \partial^{\ov\n}\vf \partial^{\m}\ov\vf\right) \; . \nn
\end{align}
\end{widetext}
Upon integrating out $f$ and $\bar{f}$ including these cubic terms one recovers the standard cubic double field theory action of \cite{Hull:2009mi}, up to field redefinitions. 
However,
in the following it will be easier to stay in the first-order formulation and impose the gauge fixing condition 
 \be\label{gauge}
   f_{\mu}=\bar{f}_{\bar{\mu}}=0\,. 
 \ee
In principle, this  leaves residual gauge transformations with $\lambda_{\mu}=\partial_{\mu}\chi$ and 
$\bar{\lambda}_{\bar{\mu}}=-\bar{\partial}_{\bar{\mu}}\chi$, but these are `trivial gauge parameters' whose action on fields vanishes. 
The above gauge  condition  is actually implied by a  well-known gauge condition for SFT, known as Siegel gauge: 
 \be
  b_0^+|\Psi\rangle = 0\;, 
 \ee
where $b_0^+= b_0+\bar{b}_0$. Indeed, acting with $b_0^+$ on (\ref{Psi}) and using $b_0^+|{\bf 0};K \rangle = 0$ one finds that the terms 
in the first and second  line are annihilated, while the terms in the third line, using $\{b_0^+, c_0^+\}=1$, give a non-vanishing contribution. 
The Siegel gauge condition thus precisely amounts to setting 
$f_{\mu}=\bar{f}_{\bar{\mu}}=0$. One may also verify that setting $x=\bar{x}$, replacing $f$ and $\bar{f}$ by their lowest-order on-shell values, 
 and  setting the scalar dilaton to zero, 
 this gauge condition implies the familiar de~Donder gauge
$\partial^{\mu} h_{\mu\nu}-\frac{1}{2}\partial_{\nu}h=0$. 

Using the gauge condition (\ref{gauge}) in (\ref{fullSFT}) one obtains 
\begin{align}\label{gaugefixedSFT}
&S_{\rm SFT}=\int d^Dx \, d^D\ov x \,\Big[ \,\frac{1}{4}e_{\m\ov\m}\square  e^{\m\ov\m} +2\ov\vf\square \vf
+ {\cal L}_{\rm DC}^{(3)} \\
&-\frac{1}{8}e_{\m\ov\n}\big(\ov\vf \partial^{\m}\ov \partial^{\ov\n}\vf+\vf \partial^{\m}\ov \partial^{\ov\n}\ov\vf
-\partial^{\m}\vf\ov \partial^{\ov\n}\ov\vf
\nn 
-\ov \partial^{\ov\n}\vf \partial^{\m}\ov\vf\big) \nn
\Big].
\end{align}
For the quadratic theory we had to integrate out the dilaton $\phi\equiv \varphi-\bar{\varphi}$ from double field theory in order to show that it equals  the double copy of Yang-Mills. 
A subtlety at this stage is that, after picking Siegel gauge, it is no longer true that only 
the combination $\phi\equiv \varphi-\bar{\varphi}$ enters the action. 
Thus, we  have to integrate out the pair of fields $(\varphi,\bar{\varphi})$. 
Since these enter the action (\ref{gaugefixedSFT}) only quadratically, at tree-level integrating them out  just amounts to setting 
$\varphi=\bar{\varphi}=0$.\footnote{Indeed, 
there are no tree-level diagrams for only external $e_{\mu\bar\mu}$ states that involve these fields  and hence setting them to zero is the correct procedure of integrating  them out 
at tree-level.} Alternatively, we could integrate out $\varphi$ and $\bar{\varphi}$ from the full action (\ref{fullSFT}) before gauge fixing, which owing 
to the linear couplings in (\ref{quadraticFirstOrder}) yields $\varphi = \square^{-1}(\partial_{\mu} f^{\mu})+\cdots$, etc.,  hence leading to a  non-local action. 
But since these non-local terms involve $f$ and $\bar{f}$ they disappear in the gauge (\ref{gauge}), and so the two operations of 
fixing a gauge and integrating out $\varphi$, $\bar{\varphi}$ commute. 
Either way, cubic double field theory in Siegel gauge precisely coincides with the double  copy of Yang-Mills theory upon integrating out the scalar fields.

\section{Summary and Outlook}

In this letter we have given   the arguably most direct  double copy prescription for the Yang-Mills action  to cubic order 
and shown that it yields double field theory to this order (in which the duality invariant dilaton has been integrated out). 
Remarkably, at quadratic order this relation holds at the level of gauge invariant actions, i.e., without the need to impose a gauge condition 
or to introduce extra fields.  While the match of the cubic terms does require gauge fixing, this gauge condition is well-known 
in  closed string field theory, from which double field theory was originally derived, as the Siegel gauge.  

The question is whether, and if so how, this relation between Yang-Mills theory and double field theory extends to higher order in fields. 
The quartic part of the Yang-Mills Lagrangian (\ref{actionYM})  can be written as 
 \be
  {\cal L}_{\rm YM}^{(4)} \ =  \ -\frac{1}{4} \,\gYM^{2}\,\kappa^{ab} f_{acd}\,  f_{bef} A^{\mu c} A^{\nu d} A_{\mu}{}^{e} A_{\nu}{}^{f}\;. 
 \ee
One subtlety in giving a double copy prescription for the quartic vertex directly is that although  
with (\ref{DCCartan}) and (\ref{eq:precripcubic}) we have given double copy substitution rules for $\kappa_{ab}$ and $f_{abc}$,  the quartic 
vertex in the above form also requires the \textit{inverse} $\kappa^{ab}$ of the Cartan-Killing metric, but  $\bar{\Pi}^{\bar{\mu}\bar{\nu}}$ is not invertible.  
Nevertheless, a natural proposal for the double copy is  a quartic Lagrangian in momentum space of the 
structural form 
 \begin{align}
  {\cal L}_{\rm DC}^{(4)} \  \propto \ & \eta^{\bar{\tau}\bar{\kappa}} \,\bar{\Pi}_{\bar{\tau}\bar{\mu}\bar{\nu}}\, \bar{\Pi}_{\bar{\kappa}\bar{\rho}\bar{\sigma}} \,
  e^{\mu\bar{\mu}} \,e^{\nu\bar{\nu}} \,e_{\mu}{}^{\bar{\rho}} \,e_{\nu}{}^{\bar{\sigma}} \nn\\ &
  + \eta^{\tau\kappa}\,\Pi_{\tau\mu\nu} \,\Pi_{\kappa\rho\sigma}\, e^{\mu\bar{\mu}}\, e^{\nu\bar{\nu}}\, e^{\rho}{}_{\bar{\mu}}\, e^{\sigma}{}_{\bar{\nu}}\,, 
 \end{align}
which would give rise to a two-derivative action.  As it stands, this proposal is incomplete since each of the $\Pi^{\mu\nu\rho}$  
depends on three momenta (c.f.~(\ref{3indexPi})), while in the above terms there are only four $e$ fields, but at least structurally one can see that such an action 
reproduces many terms of the quartic double field theory, see \cite{Boels:2015daa}.

It is clear that new ingredients are needed in order to realize double copy at the level of actions to all orders in 
fields. In particular, the perturbative expansion of double field theory is non-polynomial  and so a match with the Yang-Mills action, 
which is quartic, requires a suitable reformulation of both theories. Indeed, double copy suggests to bring both actions 
to cubic form upon introducing auxiliary fields, see e.g.~\cite{Bern:2010yg}, and more generally one may expect additional fields in order to realize the
double copy in a manner that is off-shell and  fully gauge invariant.  
This may also provide a new perspective on the important open problem, alluded to in the introduction, of constructing a weakly constrained double field theory 
beyond cubic order. While in principle such a theory can be derived from the full closed string field theory by integrating out all massive string modes 
that do not belong to the double field theory sector \cite{Sen:2016qap,Arvanitakis:2021ecw}, it would be greatly beneficial to have an efficient formulation 
of this  theory as a double copied Yang-Mills action. 

It would also be interesting to understand whether the web of double copy constructible theories, i.e.~including various
amounts of supersymmetries, gauged gravity theories and also non-gravitational theories, may be connected  
to double field theory.

\subsection*{Acknowledgements}
We have greatly benefited from discussions with Roberto Bonezzi who in particular pointed out the relevance of the Siegel gauge. 

This work is supported by the Deutsche Forschungsgemeinschaft (DFG, German Research Foundation) - Projektnummer 417533893/GRK2575 ``Rethinking Quantum Field Theory" 
and by 
the European Research Council (ERC) under the European Union's Horizon 2020 research and innovation programme (grant agreement No 771862).

\end{document}